# Low-cost Highly-interoperable Multiplatform Campus Network: Experience of YARSI University

Surya Agustian[1], Sandra Permana[2], Salman Teguh Pratista[3], Syarifu Adam[4], Iswandi[5]

[1,2]*Faculty of Information Technology, YARSI University,*
[1,2,3]*CMIS (Center for Management Information System), YARSI University,*
[1,3,4,5]*PT Optima ITT (Information & Telecommunication Technology)*
[1-5]*Jl. Letjend. Suprapto Cempaka Putih, Jakarta Pusat, 10510, Indonesia*
Tel. +62-21-4206674-76
[1]*surya.agustian@yarsi.ac.id,* [2]*sandrapermana@yahoo.com,* [3]*step2k@yahoo.com,*
[4]*adam@optimaconsultant.com,* [5]*iswandi@optimaconsultant.com*

*Abstract*—To some organizations, building campus network is sometimes considered to be very expensive; and this has made the project uneasy to perform. Moreover, if the organization without sufficient IT knowledge does not have capable IT engineers, leaving this project to third parties without supervision would lead to unexpected larger expenses. For this reason, in the year of 2003, YARSI University formed CMIS (Center for Management Information System) to perform tasks in designing, operations and maintenance of campus network and its services. By combining Open Source operating system run on a local assembled personal computer as gateway and router, and switching technology from Cisco, we designed a low-cost UTP-based campus network which covering rooms and buildings in YARSI environment. Meanwhile the internet access through several broadband connections and dedicated wireless was shared to more than 100 simultaneous users by a captive portal system. With this strategy, we can significantly reduce cost for purchasing, maintenance and operations of network infrastructure and internet access. Our model in designing low-cost campus network and internet connections could be adopted by rural community or organizations that have limited budget to have internet access.

## I. Introduction

In the internet booming in Indonesia after 1998 crisis, only a few organizations could have network infrastructure and internet access. YARSI University started to provide internet access for students in the early 2001, by opening public access services to accommodate students' need for literature search. At that time, a 64kbps leased line was used with no less than Rp. 12,000,000 subscription cost a month, a fantastic amount of budget for a very limited bandwidth when compared to the recent technologies. Besides that, this speed only served no more than 10 computer clients in a small room, while other office rooms, laboratories, and classrooms were stand-alone without networks. Recently, that amount of money could be used to purchase about 1 Mbps on a dedicated connection type like wireless, leased line, fiber optic, and some others.

In fact that internet has a strategic role in higher education, a campus network with sufficient internet bandwidth is absolutely required. A wise and smart strategy was required in designing this networking project, to balance the need and the limited budget. Open Source (OS) becomes a smart choice and cost efficient to address this problem, because there is no investment needed in purchasing OS software, and we are allowed to use the resource even in commercial sector. The booming of Facebook and Wordpress are some of the amazing success stories of web applications which were built on Open Source, that are used by millions of people in the world, and invite commercial interest. Based on this reality, some institutions of higher education included Open Source course in their curricula [1].

With empirical experience, in early development of campus network in 2003, we started by searching references of network solution in the OS domain. Based on TCP/IP routing knowledge, hand-on experiences with UNIX, and a preliminary study to Cisco switching technology, we found a cheap solution to develop campus network with limited LAN segmentations. By using VLAN concept from Cisco switched network, we divided faculties, offices, laboratories, library and public access facilities in the university environment into several LAN segments. The routing between VLANs and internet was performed by a FreeBSD server, which was installed on a PC. Meanwhile, the internet bandwidth was provided by an ISP with wireless technology.

Nowadays, some internet broadband technologies i.e. ADSL (Asymmetric Digital Subscriber Line), cable modem and mobile (3G) are sold with a relatively low subscription price. This is an opportunity for more people to access the internet with better speed. To get more advantages, we use some ADSL connections to increase total internet bandwidth for our campus network. A captive portal has been developed to distributed internet load from local LAN to internet through several proxies. Each proxy server was connected directly to customer premise equipment (e.g. ADSL modem, wireless receiver) on each ISP, servicing any web pages requested by users.

For this model, YARSI University only spends about nine million rupiahs a month for internet access subscription, with up to 7 Mbps aggregated bandwidth. A low-cost network solution was found and could be an alternative model to rural area with limited budget.

The next section will describe the strategy to build the low-cost campus network as implemented in YARSI University environment. Section III explains about the internet connection design, while the Section IV describes the latest development of our multiplatform campus network. Section V describes the network utilizations, and the last section will be conclusion of this paper.

## II. CAMPUS NETWORK INFRASTRUCTURE

Considering that the university environment is located in a relatively small area, which is about 2.5 hectare, and divided into three clusters, it is possible to design a UTP-based campus network on each cluster and several options in interconnection method between clusters. The first cluster is a 4-floor old building, which has more than 20 rooms (classrooms, labs, office) on each floor. The main cluster is a new 13-floor tower building which is still in finishing stage. The last cluster is the old buildings that will be demolished and replaced into another 13-floor tower building, planned for hospital. The optimal method we chose for interconnecting the three clusters was wireless bridge and UTP backbone, based on the distance and analysis that only internet packets on port 80 were found and needed by the users in remote clusters. Besides, it is temporary until the allocated rooms in the new 13-floor building for the users are finished and ready to use. Once they all move into the new building, the old one will be demolished.

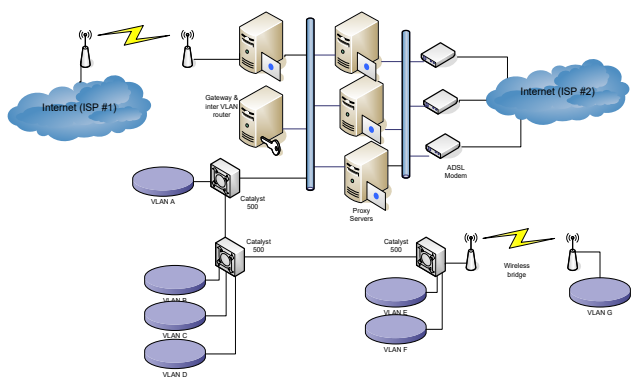

Figure 1. Low-cost campus network design

Figure 1 describes the scheme of our campus network design, and it has already proven to be reliable, scalable, and low-cost implementation. LAN segmentation is formed by VLAN encapsulation on Cisco Catalyst Express 500, which is routed by a UNIX FreeBSD server. The server also acts as firewall and captive portal to redirect internet packet request from users to proxy servers. With this configuration, we can reduce cost for backbone and active devices, which in some organizations will cost too much. The need of distribution layer switch that was usually provided by higher Catalyst series i.e. Cat3600, was replaced by an entry-level manageable switch from Cisco, Catalyst Express series, which is usually used as access layer switch. The DHCP services, VLAN management, inter VLAN routing and access control list tasks are performed by PC server. This combination (Catalyst Express and FreeBSD) has replaced the function of communication server that usually was performed by distribution layer switch, and the purchasing budget is reduced to three times cheaper than conventional configuration, which is usually proposed by certified professional network consultant.

From server side, only a trunk backbone cable is connected to its network interface. This connection is treated as trunk where all VLAN segments transmit and receive data. Any data frames entering the interface are analyzed, and then forwarded to the destination network or blocked. By default, all the traffic to outside network is blocked. Only the interested packet will be forwarded to the destination based on access list rules.

For security reasons, we will describe a modified IP addressing scheme for the proposed campus network, and eliminate the DMZ (demilitarized zone) from network diagram in Figure 1, which covers server farm (web, database, mail, etc). The management VLAN has one segment IP address 192.168.1.0/24, where the trunk cable from each catalyst and proxy servers are plugged. Captive Portal Server or gateway has virtual VLAN interfaces configured on the same gigabit network interface. Each trunk cable entering server interface or Cisco Catalyst Express 500 series uplinks (2 ports are gigabit Ethernet port) are cat6, allowing maximum speed access up to 1000 Mbps. With this setting, the preliminary requirement of backbone is fulfilled.

Other catalyst switch ports are configured as VLAN tagging mode, and connected to unmanageable 100 Mbps switches as access layer. At the end point, users' PCs are connected to those switches with certain VLAN IP addresses acquired from DHCP server. As catalyst switch does not support layer 3 encapsulation, DHCP service is provided by gateway server. We configured other VLANs with different IP address, e.g. VLAN 2 is 192.168.2.0/24, 192.168.3.0/24 for VLAN 3, and so on. It is also allowing different users' location to have flexible VLAN IP address by configuring the appropriate switch port, when they are moved to another room or building.

## III. INTERNET CONNECTION DESIGN

To figure out the users' need, we analyze the data traffic across VLAN gateway. Most of packets are http, and some of them are well known port number which is usually used by IT staff. Since we did not implement active directory yet, virus like traffic i.e. netbios family, are dropped by access list. We only passed some selected well known ports in the access lists, and some other ports based on user's trouble reports. This configuration will prevent each VLAN segment to infect others while a computer attacked by viruses or spyware.

What we have done was in line with Cisco strategy in developing new framework of campus network evolutions. It is said that the campus network needs to target services to end-user as the new business focus, i.e. applications, information, and network security move rapidly to address end-user needs and interest [2].

Since only web request as the most end-users' interest which would be considered as the interesting packet to be forwarded by ACL, we decided to choose the best connection method for optimal performance and cost. Table 1 shows the comparative price on several internet technologies. From the table, we could map the requirement, availability and budget for our gateway to internet i.e. several broadband ADSL connections which have a reasonable price for small to medium business.

To prevent single point of failure, we also subscribe a dedicated wireless connection of other ISP. In 2005, the total incoming bandwidth we had possessed was approximately 1,256 kbps (a 512 kbps wireless and two 384 kbps ADSL) which cost about five millions rupiahs a month. This cost was less than a quarter compared to a dedicated 1 Mbps broadband wireless or leased line at that time. Recently, broadband ADSL technology became cheaper and bandwidth increased. With the wide range of copper telecommunication infrastructure provided and installed by PT Telkom that reaches more rural area in Indonesia, the internet penetration will grow faster using ADSL technology.

Table I. The Comparison of Several Internet Connection Methods (offered in 2008-2009)[1]

| Technology | Total cost/month | Availability |
|---|---|---|
| ISP A: broadband SOHO (ADSL) up to 1 Mbps | 2,000,000 IDR | Yes |
| ISP A: dedicated wireless 256 kbps 1:1 | 6,000,000 IDR | Yes |
| ISP B: dedicated wireless 128 kbps 1:1 | 5.000.000 IDR | Yes |
| ISP B: dedicated wireless 256 kbps 1:1 | 6,500,000 IDR | Yes |
| ISP C: broadband ADSL Up to 1 Mbps | 800.000 IDR | Yes |
| ISP D: broadband FO 1 Mbps 1:1 | 34,500,000 IDR | No |
| ISP D: broadband wireless 1 Mbps 1:1 | 30,500,000 IDR | Yes |
| ISP D: broadband FO 256 kbps 1:1 | 18.000.000 IDR | No |
| ISP D: broadband wireless 256 kbps 1:1 | 11,000,000 IDR | Yes |
| ISP E: broadband wireless up to 512 kbps 1:4 | 2,000,000 IDR | Yes |

[1] collected from offering letters and quotations that came to author

*Captive Portal*

For more details on how a user can access a URL in the proposed network campus, we can refer to the Figure 2 below. A packet requested for port 80 or 8080 will be captured by VLAN gateway, which acts as captive portal. This server forwards the request to an authentication page, asking for username and password. In the early development, user-name and password were stored in a MySQL database. Once a user was authenticated, other information like bandwidth utilization, IP address, login time, and activities are stored in database. When idle for more than 10 minutes, the user would automatically sign off from the system.

A php script ran as background process to investigate all incoming traffic from VLANs. This script managed forwarding rules in the ACL, activities logging, and real time bandwidth utilization of each users. A set of implicit rules were defined to balance the load among proxies. Simple Network Management Protocol (SNMP) was used to get information of interface status (up or down) and current traffic passed the interface in real time mode. The purpose was to simplify network administrator tasks in controlling and monitoring the network performance.

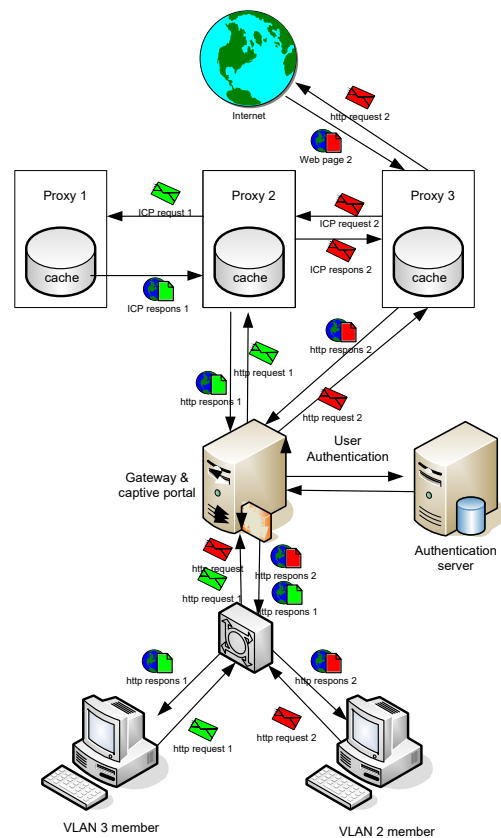

Figure 2. The description of http data across network

*Proxy Load Balancing*

Other developed feature of the gateway is proxy load balancing. A load balancing scheme is designed to distribute the internet traffic load, by analyzing layer 4 data across VLAN interfaces on the server. The packets to port 80 or 8080 (HTTP request) are forwarded to a certain proxy server based on load balancing rules. Then proxy server will check its cache for local copy of a requested web page, and if exists, an HTTP response which contains requested web page is passed to the client. Otherwise, the proxy on behalf of the client retrieves the requested page to internet. This method will improve the performance and the response time in user's perspective.

Further, several proxy servers could be configured to be a local synergy and use WCCP (Web Cache Communication Protocol) to communicate to router, which was developed by Cisco [3], [4] or use CARP (Cache Array Routing Protocol) developed by Microsoft [4], [5]. Other collaboration methods of communication use ICP (Internet Cache Protocol), cache digest, and other cooperation protocols [6]. ICP is said to be the best example to produce global synergy of proxies within internet across different networks [4].

For that reasons, ICP becomes the popular and typical method used by various organizations when implementing collaborative internet caching system. Several proxies are configured to work together within ICP environment with two choices, equal sibling or parent-child configuration. A proxy will retrieve from its cache any incoming HTTP request, and if not found, it will send ICP request to sibling in the network for the requested page. When no proxy in the network has the page in their cache, the origin proxy will retrieve the page directly to remote server. Parent-child configuration works similarly, if a requested page comes to a child proxy, it first searches for local copy. Whenever not found, the child requests the page to its parent. If parent proxy does not have either, it will get the pages to internet. Only the parent has responsibility to retrieve any un-cached pages directly from remote servers, and the child gets from its parent cache [4], [5], [7].

*Local Synergy Design*

Proxy family is placed in a separate LAN segment in the campus network. Referring to Figure 1, each proxy has two network interfaces, but none of them are configured to allow IP forwarding as a router. This is aimed for security reasons, to reduce vulnerabilities in many redundant internet connections.

The first interface of each proxy is connected to server segmentation (core link), where various busy network traffic passes through. Another interface is connected to proxy segmentation (proxy cluster), where the ICP and HTTP data flow among interfaces. We did not design a cross connection between proxy and ADSL modem to avoid single point of failure. By forming a separated LAN segment for proxies and ADSL modems, an alternate route could be set to a certain proxy when a problem occurred to its default gateway. SNMP could be utilized for early detection in internet connection, and a simple shell script could be written to automatically alternate the default route to other ADSL modem.

If all broadband connections failed, captive portal will forward the http request from clients to the proxy which connected to dedicated wireless. With this strategy, we can reduce a possibility of single point of failure, and users are always able to access internet, unless all connections are in trouble.

Considering that only simple tasks will be performed by proxies, we utilized old PCs (Pentium III or Pentium IV) with limited specification to be a part of the proxy system. With this strategy, we can save more budgets for internet infrastructure, and help to utilize unused old PCs which are still in good conditions.

*Network Address Translation* (NAT)

To accommodate users' need to access outside network other than web pages, Network Address Translation (NAT) is set up. Basically, all clients in VLANs are permitted to access other applications outside network as long as they are reported to Network Administrator. When the applications are blocked, net-admin analyzes the traffic, determines the blocked port, and then creates new access lists to permit the applications.

For these requirements, another PC is used as router and firewall, which is not drawn in Figure 1. This router will allow access to university information resources (official websites and corporate mail), both from inside campus network and from internet.

## IV. LATEST DEVELOPMENT

To form a better security and easier user management, we implement single sign on for many applications, e.g. internet access authentication, email account, information systems, e-learning, blogs registration, and profile management. We implement active directory with LDAP which runs on a Linux server. Clients' PC with Windows operating system are joined to domain controller, which provided by Linux PDC (Primary Domain Controller). A Backup Domain Controller is also configured on Linux. By logging on to domain, users could store their personal files in their home directory on file server, and easily access them anywhere in the campus network.

Internet authentication is also served by LDAP. It is performed by forwarding any http request to YARSI website where users have to log in. The aim is to force users read something in the university official website, as it is common that many members of the organization are not familiar with their organization website.

In early development, captive portal forwards authentication request to database server, but today, the username, password, and other personal information are stored in LDAP. Authentication data is sent by a secure http connection after log in to YARSI website. Database is still used to store current log activities of the users, IP addresses from where they access internet, log on time, size of data transfer and bandwidth utilization.

Mail and other information system are also LDAP

authenticated. All the mentioned services are provided by several PC servers that placed in the core segment in a VLAN. As this segment becomes the busiest LAN in the campus networks, we used catalyst 2950 series. In the future development, we will replace this core switch with higher level catalyst, i.e. layer 3 Catalyst 3560 series with 24 gigabit Ethernet port.

Our campus network then become sophisticated, that with several operating systems (i.e. FreeBSD and Linux as server, Windows XP and Vista as clients), several Open Source applications (i.e. web, database, LDAP, domain controller, firewall and proxy), several internet connections (dedicated wireless and broadband ADSL links), and layer 2 and layer 3 switching technology from Cisco. This synergy produces a low-cost highly-interoperable campus network that has fulfilled end-users' needs with complete services.

*Network Performance*

Currently, internet traffic growth has increased rapidly. We count the intranet users who access internet and remote users who browse university website. The result is shown in YARSI website. The maximum count on working days reached 200 users from outside network and 120 concurrent users from intranet. Figure 3 below shows the capture of real time count of user who access YARSI website on May 15, 2009 at 13:10 Jakarta time. The number of Civitas online represents the real time users who access internet after authenticated by LDAP.

many users are concurrently accessing the internet, the proxies system still can handle. It is shown by the real time traffic from each SNMP capture, which no figure reaches the peak.

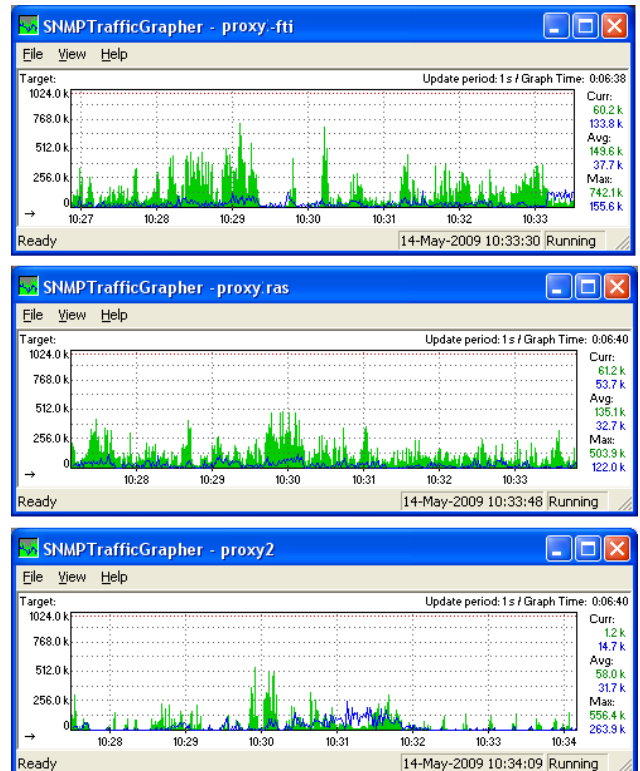

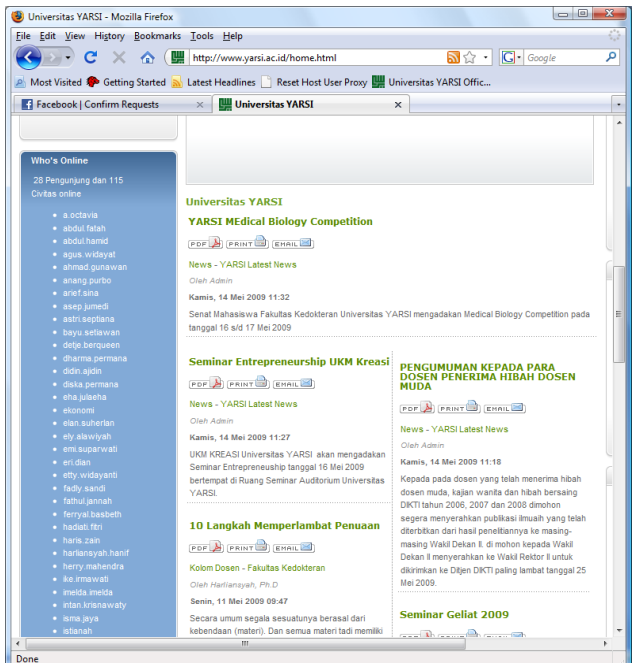

Figure 3. Online authenticated users in YARSI website

Figure 4. Some captures of proxies' load

A general issue in low-cost Open Source is the contribution to widespread sophisticated development practice [8]. We realize that our low-cost campus network model with high interoperability components, devices and platform, makes the system quite complex and not easy to hand over. For this issue, an organization who wants to implement this model has to have a capable network engineer to design and maintain the operations of the overall system.

## V. CONCLUSION

Theoretically and having been proven, our network campus model could provide the need of accessing documents from internet in a rational tolerance delay perceived by user; on the other hand, organization is not burdened with the expensive cost for purchasing and operations. In six-year experience, implementing this strategy has provided great advantages in terms of cost saving, and could be an optimal solution for small and medium business with limited internet budget.

## VII. ACKNOWLEDGMENT

The author would like to thank Prof. Dr. Jurnalis Uddin, PAK, the founder of Yayasan YARSI, for giving the opportunity, confidence and support to the extraordinary IT development team, from CMIS period to Optima ITT, to do designing, research and experiment on

Figure 4 shows the proxies' traffic loads which were captured on business day, Thursday 14 May 2009, on about 10.30 Jakarta time. Referring to Figure 4, implementing local synergy with several proxies and redundant internet connection as explained in Section III, can reduce the internet traffic load significantly. Although

YARSI campus network and its services. Author also would like to thank Mrs. Novi Rahayu for reading the article and giving some advice on the language.